# Beyond the Golden Record: Toward a Design Theory for Trustworthy Master Data Management with Self-Sovereign Identity

Niklas Schulte[1, 3(✉)], Isaac Henderson Johnson Jeyakumar[2], Michael Kubach[2] and Christian Janiesch[1,3]

[1] Fraunhofer ISST, Dortmund, Germany
niklas.schulte@isst.fraunhofer.de
[2] Fraunhofer IAO, Stuttgart, Germany
{isaac-henderson.johnson-jeyakumar, michael.kubach}@iao.fraunhofer.de
[3] TU Dortmund University, Dortmund, Germany
christian.janiesch@tu-dortmund.de

**Abstract.** Ensuring the timeliness and reliability of master data remains a persistent challenge for many organizations. To mitigate these quality deficits, organizations frequently rely on commercial data brokers. However, this practice creates strategic dependencies and poses significant business risks, particularly as providers typically disclaim liability for the accuracy of the supplied data. In contrast, modern data ecosystems enable the trusted sharing of data assets with strong data sovereignty. In this paper, we address this paradigm shift by deriving a nascent design theory for trustworthy master data management based on self-sovereign identity. The theory is grounded through a hermeneutic literature review combined with industry expert interviews and instantiated through integration into a reference architecture for data spaces. Following an evaluation through additional industry expert interviews, our work provides a framework for a trustworthy master data management in data ecosystems that is reliable, sovereign, and accountable.

**Keywords:** Master Data Management, Self-Sovereign Identity, Data Ecosystem, Data Space, Data Sovereignty.

## 1 Introduction

The industrial landscape is currently undergoing a paradigm shift towards collaborative, cross-sector data sharing initiatives such as Gaia-X [1] and Catena-X [2]. In these socio-technical data ecosystems, stakeholders are moving away from centralized data platforms toward sovereign data sharing, leveraging interoperable, decentralized architectures and data usage control mechanisms. However, current Master Data Management (MDM) practices are not adhering to this paradigm shift. Organizations often rely on commercial data brokers to improve their master data quality [3, 4], which introduces strategic dependencies and liability risks, as these data brokers are structurally unable to guarantee the validity or timeliness of the supplied data[1]. The business consequences are significant. Erroneous master data directly causes supply chain disruptions, failed regulatory reporting, and costly manual reconciliation efforts [5–7]. Moreover, as regulations such as GDPR, the Data Act, and eIDAS 2.0 impose stricter requirements on data provenance and accountability, reliance on unverifiable third-party master data increasingly constitutes a compliance risk. This practice places organizations in a precarious position, as they remain dependent on externally supplied master data whose validity is unverifiable, and no provider accountability can be assumed. In contrast to this, modern data ecosystems allow the accuracy, timeliness, and accountability of data to be attested directly by the data providers themselves, without any central intermediaries. Yet, this transition presents a critical challenge for current MDM practices, on which companies rely for regulatory compliance, process integration, and business relations [3, 6, 8].

Thus, in this paper, we explore the potential of Self-Sovereign Identity (SSI) for a decentralized MDM approach. While SSI is widely recognized as a trust-enhancing technology in modern data sharing environments, its application has been largely restricted to providing evidence for access control or compliance purposes, but not for the management and exchange of master data records [9–11]. We identified a significant research gap between the two domains: traditional MDM does not account for sovereign, decentralized data exchange, while current SSI applications lack the semantic rigor and integration capabilities required for the processing of master data records. To this end, we apply a Design Science Research (DSR) approach to derive a nascent design theory for trustworthy MDM with SSI. We address the following research question: *How can cross-organizational master data management be designed to ensure trustworthiness in decentralized data sharing environments?*

The remainder of this paper is structured as follows: Section 2 provides the background on MDM, data ecosystems, and SSI, followed by our DSR methodology in Section 3. Section 4 grounds the research through a literature review and expert interviews, which inform the design principles and features derived in Section 5. We demonstrate the theory's applicability in Section 6 and evaluate it via expert interviews in Section 7. Lastly, Sections 8 and 9 present our discussion and conclusion, reflecting on the implications of our design theory for MDM practice and research, as well as its limitations and areas for future work.

## 2 Research Background

Our research is positioned at the interface between the traditional discipline of MDM and the emerging field of decentralized data ecosystems, which typically leverage the paradigm of SSI to establish digital trust between participants. Although existing research addresses solutions for decentralized sharing, the implications of these technologies for current MDM practices remain underexplored. Specifically, the shift away from a reliance on central data brokers for data quality improvements and the move from centralized data governance of "Golden Records" towards distributed models represents a significant architectural challenge that has yet to be fully reconciled with the principles of data sovereignty.

---

[1] This is reflected in the terms of use of commercial providers such as North Data GmbH and Dun & Bradstreet, which offer data correction mechanisms rather than guarantees of accuracy [6, 7].

### 2.1 Master Data Management and the "Golden Record"

Traditionally, MDM comprises methods and tools to ensure high data quality and reusability across organizations' business processes [12]. The central objective of traditional MDM is the establishment of a "Golden Record" to guarantee a single, authoritative source of truth that consolidates duplicate or conflicting data records into a unique, reliable representation, defined as "analytical MDM" by Otto et al [3]. Insufficient master data management leads to inaccurate and incomplete information, which results in flawed stakeholder decision-making and potential financial losses for organizations [13].

In practice, master data is often collected company-internal in centralized architectures. As described by Loshin [14], MDM relies on central hubs or registries to harmonize data from various sources across organizations. This reliance on centralized systems extends beyond internal systems to modern cross-organizational data ecosystems. A prominent example of a central infrastructure for MDM is the Cofinity-X Golden Record Service (GRS) [15]. Currently operating in the Catena-X ecosystem, the GRS functions as a centralized intermediary that aggregates business partner data to perform duplicate management, error correction, and manual data enrichment. While this is an effective way to establish data quality, such centralized master data aggregations pose risks to data breaches and create unwanted central dependencies in decentralized data-sharing environments.

### 2.2 Self-Sovereign Identities and Data Ecosystems

Self-Sovereign Identity (SSI) represents a paradigm shift in digital identity management, away from centralized control and governance of identity information towards the creation of a common identity layer for the internet [16]. The goal of SSI is to give users an independent digital existence and full control over their digital identities [16, 17]. Following the W3C recommendation for Verifiable Credentials [18], SSI mainly involves three roles: issuers, holders, and verifiers. Issuers issue identity claims as Verifiable Credentials (VCs), holders manage these credentials in digital wallets, and verifiers request and validate presented credentials from holders. Its adoption promises privacy, data security, and trust in digital ecosystems [10].

In data ecosystems, SSI-based digital identity management serves as a critical element for facilitating secure and sovereign data sharing across organizations [19]. For example, Gaia-X uses SSI concepts to establish trust not only in organizations and people, but also in services, data assets, and technical components, which receive verifiable self- descriptions linked to their identities. These self-descriptions, expressed as VCs or attestations, enable dynamic trust assessment across federated ecosystems. Although sovereign data exchange should ideally occur directly between two participants, current data space implementations often rely on intermediaries like clearing houses or brokers to establish initial trust and mediate trustworthy transactions [20].

Consequently, the move toward self-sovereign data exchange necessitates a strategic alignment with intra-organizational MDM practices. While SSI integrations in data ecosystems manage external trust using globally unique Decentralized Identifiers (DIDs), MDM systems are required to map these external identities to internal "Golden Records". By leveraging this approach, organizations can treat data ecosystems as a high-trust source system for inter-organizational master data exchange. Hence, incoming VCs containing validated business partner data can automatically update internal master data records. Following this approach, we resolve a disconnect between external data sovereignty and internal master data processing, ensuring that the trust established in the data ecosystem directly translates to automated master data quality maintenance.

## 3 Research Method

We ground our methodological approach in the DSR process model defined by Peffers et al. [21] as illustrated in Fig. 1. To ensure the accumulation of design knowledge, we adhere to the guidelines of vom Brocke et al. [22] regarding grounding, positioning, aligning, and advancing our contribution within the existing knowledge base.

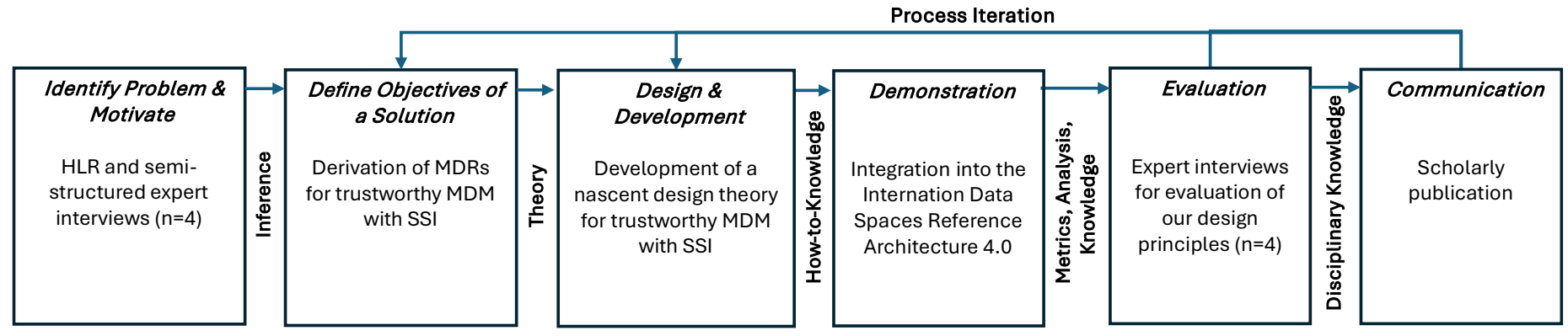

**Fig. 1.** DSR process for theory development based on Peffers et al. [21].

We employed a mixed-methods approach, bridging rigor and relevance cycles as defined by Hevner et al. [23] for problem identification. We first conducted a Hermeneutic Literature Review (HLR) following Boell and Cecez-Kecmanovic [24], an approach that treats the literature review as alternating cycles of "search and acquisition" and "analysis and interpretation." Through this, we ensure scientific rigor by grounding our work in established theories. To ensure practical relevance, we enriched these theoretical findings with semi-structured expert interviews to derive robust design requirements. We consider our problem identification as a holistic hermeneutic understanding process, combining literature review and interviews.

Based on this problem understanding, we derive a nascent information systems design theory as our primary DSR artifact. To systematically develop our design theory, we follow the hierarchical derivation approach of Herm et al. [25]. This method structures the development process into four logical levels: Meta Design Requirements (MDRs), Design Requirements (DRs), Design Principles (DPs), and Design Features (DFs). This hierarchy ensures traceability from abstract design requirements to concrete technical features. At first, the "purpose and scope" component of Gregor and Jones [26] is operationalized by defining MDRs. These MDRs delineate the class of goals our design theory addresses. Subsequently, we refine these MDRs into concrete DRs and, ultimately, into actionable “design principles of form and function”. To ensure these DPs are actionable and effective, we follow the approach of Chandra et al. [27] and Gregor et al. [28] to provide a clear rationale for each principle. Lastly, we provide a mapping from our DPs to DFs to visualize how our DPs can be instantiated by concrete technologies [25].

Continuing our DSR approach, we developed a reference architecture showcasing an integration of decentralized MDM into the International Data Space Association (IDSA) Reference Architecture Model 4.0 [29] based on the derived DPs. Through this, we demonstrate the technical feasibility of our design theory and illustrate how it can be translated into a coherent system architecture. Subsequently, we evaluate our design theory by following the Framework for Evaluation in Design Science (FEDS) proposed by Venable et al. [30]. Given the abstract nature of the design artifact, we performed an ex-ante evaluation following the “Technical Risk & Efficacy Strategy” of FEDS. This evaluation strategy allows us to assess the efficacy and logical completeness of our DPs in an artificial setting. Hence, we conducted semi-structured expert interviews to validate our DPs. Following the approach of Janiesch et al. [31], we apply the metrics of “ontological expressiveness” to guide our interviews for evaluation. This formative evaluation strategy maximizes learning and improvement by directly identifying practical barriers to applying the DPs.

# 4 Problem Identification

Following Boell and Cecez-Kecmanovic [24], our HLR proceeded in iterative cycles across three databases, Scopus, ACM Digital Library, and AISeL, supplemented by forward and backward citation search. In Iteration 1, we conducted a broad search for MDM literature on ACM Digital Library and AIS eLibrary with the query (*"master data")*. For Scopus, we restricted our search term further by adding *AND ("governance" OR "data quality" OR "golden record")* to keep the search results manageable. After screening titles and abstracts for relevance to organizational MDM practices, 19 papers were retained after filtering, enriched by 5 papers through forward and backward citation search, resulting in a total of 24 papers. In Iteration 2, we extended the search to the use of SSI and MDM in decentralized data sharing environments using the query (*"master data" OR "self-sovereign identity”) AND ("data space" OR "data ecosystem"*). Across all three databases, this yielded 27 results, of which 10 were retained after applying the same relevance criterion, enriched by 3 papers through citation search, leading to a total of 13 papers. Table 1 visualizes the paper selection process.

**Table 1**. HLR paper selection process.

| I# | Scopus | ACM DL | AISeL | Filtering and Selection | Citation Search | Σ |
|---|---|---|---|---|---|---|
| I1 | 262 | 53 | 36 | -332 | + 5 | 24 |
| I2 | 22 | 2 | 3 | -17 | +3 | 13 |

### 4.1 Iteration 1: Traditional Master Data Management

Traditional MDM typically aims towards a centralized collection of data to provide a "single source of truth" within organizations. To this end, Raharjo et al. [32] identified technical solutions to support master data collection and proper data governance as critical success factors for MDM. However, Silvola et al. [33] note that organizations frequently suffer from inadequate data ownership and incoherent data management processes. Vilminko-Heikkinen and Pekkola [12] demonstrate that successful MDM requires strictly defined data owners with explicit authorities to make binding decisions. Following Keith and Seymour [6], this requirement is difficult to enforce in practice, as organizations constantly struggle with "data silos" and "fragmented infrastructures."

To address these governance challenges, Otto and Ofner [3] include the definition of a context-dependent "Nucleus" and the integration of commercial master data providers. Buffenoir and Bourdon [34] characterize this centralized governance model as the "Panopticon Paradigm", where data quality is enforced through centralized visibility and disciplinary power of the overarching organization. Technically, this is typically realized through central registries designed to support transparent access to a unique representation of master data [14].

Despite these frameworks, centralized models face inherent limitations in cross-organizational settings. Otto and Ofner [3] further identify a "lack of downstream visibility", indicating that central authorities cannot oversee how data is used at the edge. Although Otto et al. [35] acknowledge that there is a need for external communication for MDM, their reference model treats it as an output of the internal quality processes, assuming external partners will inherently trust it. However, in fully decentralized data sharing environments, inherent trust cannot be assumed, and the authoritative power to enforce it is omitted.

### 4.2 Iteration 2: Master Data Management in Data Ecosystems

Considering modern, decentralized data sharing solutions for master data exchange, the focus shifts from company-internal data quality optimizations towards the management of identities, interoperability, and trust relationships across organizational boundaries. In contrast to traditional MDM, the collaborative nature of data ecosystems requires mechanisms for unique identification and verification without a central intermediary.

A central hindrance to data sharing in data ecosystems is the heterogeneity of data sources. Datta et al. [36] recommend a reference architecture that harmonizes heterogeneous data sources to enforce rule-based compliance. However, Altendeitering and Guggenberger [37] emphasize that this is particularly critical for relational master data records, which serve as the backbone of business processes and require dedicated quality analysis tools. To enable a clear understanding across boundaries, semantic models are inevitable. Bader et al. [38] demonstrate how semantic models can be used to define not only data resources, but also the actors involved and their connectors in a machine-readable format. Furthermore, Azkan et al. [39] emphasize that defining master data attributes requires a clear classification of ecosystem roles to determine value creation.

A further fundamental prerequisite for decentralized MDM is the establishment of trustworthy identities. Gelhaar and Otto [40] identify trust building as a primary cooperative challenge, arguing that it has direct implications for managing partner relationships. To address this, Babel et al. [10] describe the leveraging of SSI for data ecosystems. In addition, Barclay et al. [41] demonstrate how this concept extends to data accountability: they propose using VCs to create cryptographic “Bills of Materials“ (BOMs). These BOMs provide a transparent, tamper-proof supply chain record, allowing consumers to scrutinize the origin and integrity of shared data assets.

### 4.3 Iteration 3: Stakeholder Interviews

To complement our theoretical findings with practical perspectives, we conducted four semi-structured expert interviews with practitioners engaged in data space implementations and SSI infrastructure development. We employed a purposive sampling strategy, selecting participants across the primary stakeholder roles of data space ecosystems: infrastructure providers, governance bodies, and domain-specific end-users. Selection criteria were: (1) active involvement in a multi-organizational data exchange project, (2) a minimum of 2 years' experience in either SSI or MDM, and (3) decision-making authority within their respective project (Architect, Board Member, or Product Owner). The purposive selection ensures coverage across these primary roles, and in alignment with DSR methodology, the interviews serve to inform design requirements rather than achieve statistical generalizability [21, 23]. Theoretical saturation was considered reached as no new themes emerged after the fourth interview.

Participants represented diverse organizational contexts: a chief architect for SSI solutions at a major enterprise software provider (P1), a board member of a European identity cooperative (P2), a research associate working on healthcare data spaces (P3), and a product owner for MDM in the automotive domain (P4), ensuring broad coverage across technical implementation, identity infrastructure, and domain-specific MDM applications. Table 2 provides further demographics. We prepared an interview guideline with 14 questions across 6 categories derived from the HLR findings, opening each interview with a shared understanding of MDM before exploring its significance for data spaces. All interviews were recorded, transcribed, and analyzed by the authors.

**Table 2**. Interview demographics for DSR problem identification.

| P# | Role | Domain | Background in SSI? | Experience | Duration |
|---|---|---|---|---|---|
| P1 | Chief Software Architect | IT Consulting | yes | 22 years | 22min |
| P2 | Digital Identity Expert | Digital Identity | yes | 15 years | 47min |
| P3 | Research Associate | Healthcare | no | 4 years | 27min |
| P4 | Product Owner MDM | Automotive | no | 2 years | 42min |
| **Σ** | | | | | 138 min |

A recurring theme across all interviews was the challenge of maintaining current and accurate master data. P2 cited a current real-world example for this. Organizations maintain extensive datasets of business partner information, but often lack mechanisms to verify whether this data remains valid over time. For example, when contact people leave organizations without notification, often no updates are reported, which creates significant compliance risks. P2 further noted that changes to official registries can take multiple months to process, creating problematic gaps where outgoing executives retain signing authority while incoming leadership cannot exercise legitimate authority. Current approaches using commercial data brokers were identified as fundamentally problematic from a liability perspective.

P3 emphasized that services aggregating publicly available information explicitly disclaim responsibility for data validity, creating unacceptable risk exposure for business-critical decisions. P3 also introduced a different perspective, especially interesting for sensitive data sharing contexts, noting that Verifiable Presentations (VPs) enable answering specific queries without disclosing underlying data, which is essential for healthcare applications where privacy and confidentiality must be preserved. On semantic interoperability, P3 offered that rather than enforcing a single uniform data model, translation tables between established formats may prove more achievable, acknowledging that global consensus on standards in the healthcare domain remains elusive.

P4 emphasized the importance of system activation. Despite backing from major manufacturers and explicit membership targets, adoption in Catena-X remained below expectations. In addition, P2 highlighted significant potential for SSI in master data governance, describing, for example, bank account verification through VCs could eliminate multi-person approval processes currently required to prevent fraud, noting "the efficiency gain is enormous."

The expert interviews surfaced several requirements complementing our literature-derived findings. Inter-organizational master data solutions must accommodate temporal gaps between operational efficiency and official registry updates. Accountability for master data should be technically enforced through cryptographic signatures and ideally directly attested by the data owner. Identity architectures must support hierarchical relationships reflecting actual business structures. Lastly, practical adoption depends critically on demonstrating clear value propositions of the system, justifying initial investment for data ecosystem participation.

# 5 Design Theory for Decentralized MDM with SSI

In the following section, we present our nascent design theory for decentralized MDM with SSI. Fig. 2 provides an overview of the derived theory, including refinements made following our evaluation phase. While DIDs and VCs are established building blocks, their systematic composition for MDM-specific requirements constitutes the novel contribution of this work. The following Meta Design Requirements (MDRs) are derived directly from the deficits of centralized MDM identified in HLR Iteration 1 and the specific interoperability and accountability challenges revealed in Iterations 2 and 3.

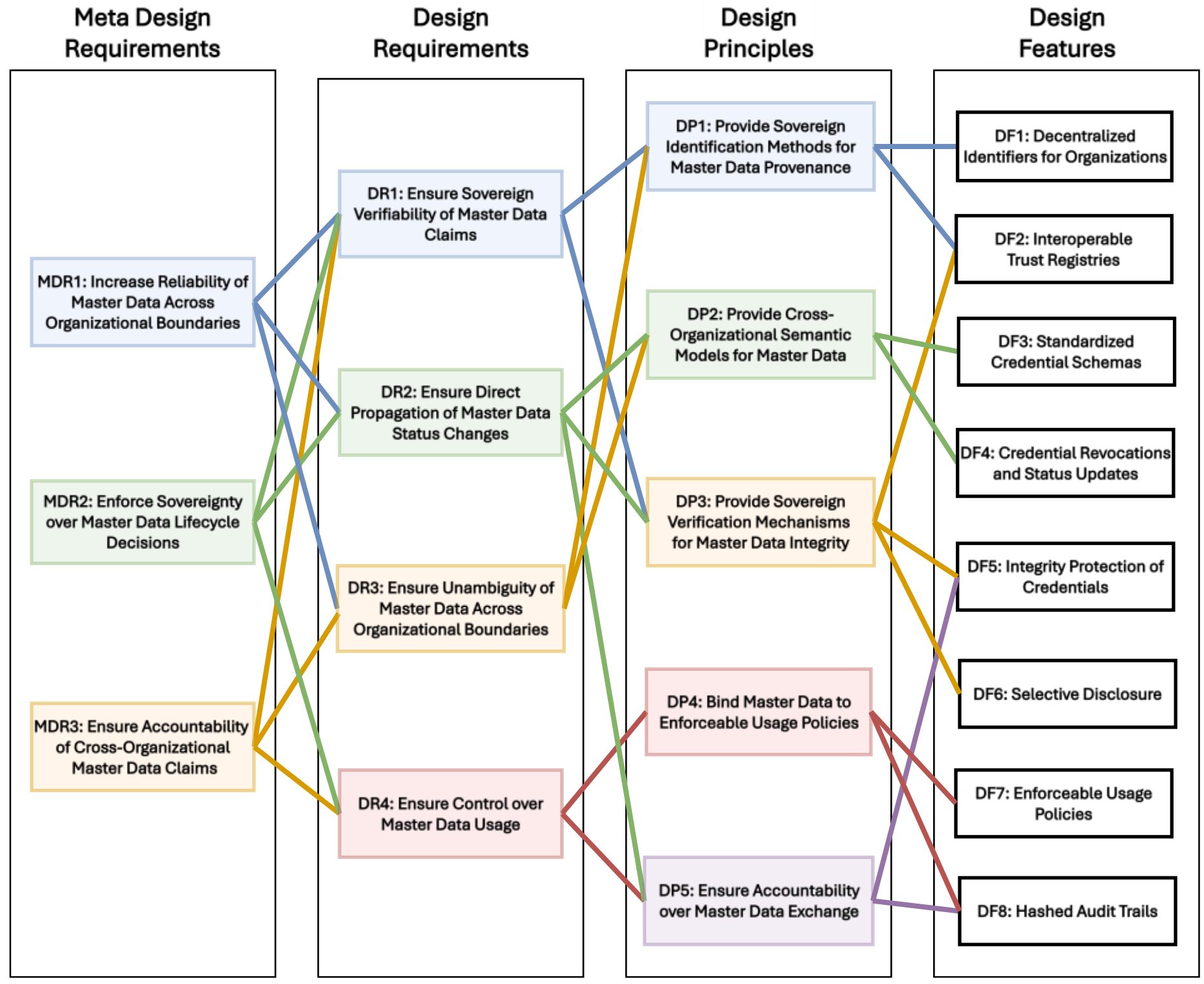

**Fig. 2.** Design theory for trustworthy MDM with SSI.

**MDR1: Increase Reliability of Master Data Across Organizational Boundaries.** Uncertainty about the status of master data is a major issue for organizations in practice. Often this is mitigated through the purchase of data from commercial data brokers [3]. While traditional MDM focuses on creating an organization's internal "Golden Record," data ecosystems require mechanisms that ensure reliability beyond organizational borders. Addressing the problematic missing updates, identified in our expert interviews (Section 4.3), MDR1 captures that inter-organizational MDM approaches need to have a similar reliability as organization-internal "Golden Records" [35]. This reliability needs to ensure the correctness, timeliness, and verifiability of master data originating from external organizations.

**MDR2: Enforce Sovereignty over Master Data Lifecycle Decisions.** In cross-organizational data ecosystems, master data assets are frequently created, shared, updated, and revoked. To overcome the "Panopticon Paradigm" of centralized governance identified in Iteration 1 (4.1) and leverage the decentralized control mechanisms of data spaces found in Iteration 2, data providers require sovereignty over master data lifecycle decisions such as issuance, modification, revocation, and usage even after data has been shared with external organizations [42, 43]. This is particularly critical for master data, as non-propagated changes can propagate inconsistencies across data ecosystems.

Therefore, MDR2 captures inter-organizational MDM approaches that must preserve the organization's sovereignty over master data lifecycle decisions.

**MDR3: Ensure Accountability of Cross-Organizational Master Data Claims.** When master data is shared across organizations, the receiving organization must be able to rely on it. Prior research on data governance emphasizes that effective inter-organizational data sharing requires clear accountability structures that allow actors to attribute data-related decisions and outcomes to identifiable parties [12]. In the context of MDM, accountability is particularly critical because master data often serves as a foundation for regulatory compliance, contractual relationships, and operational decision-making across organizational boundaries [35]. Directly responding to the lack of liability assumed by commercial data brokers identified in Iteration 3 (Section 4.3) and the "lack of downstream visibility" noted in Iteration 1 (Section 4.1), the receiving organization must be able to rely on shared data. Therefore, MDR3 states that inter-organizational MDM approaches must ensure accountability of master data claims to determine who issued, updated, or revoked specific master data.

On this basis, we operationalize the MDRs by deriving more granular DRs, which provide specifications that facilitate the practical instantiation of the design [25]. These DRs are subsequently used as the foundation to derive DPs.

**DR1: Ensure Sovereign Verifiability of Master Data Claims.** In contrast to centralized authorities that guarantee reliability in traditional MDM environments, for decentralized environments, we require the data itself to be verifiable. Drawing from the requirements of reliability (MDR1), sovereignty over lifecycle decisions (MDR2), and accountability for claims (MDR3). DR1 aims to improve trust between organizations in data ecosystems through direct verifiability of master data records as a key component for sovereign and cross-organizational data sharing.

**DR2: Ensure Sovereign Direct Propagation of Master Data Status Changes.** Master data frequently gets outdated or potentially even needs to be revoked by the data owner. These lifecycle decisions should be propagated in a sovereign way (MDR2) and thus increase the reliability of master data across organizational boundaries (MDR1). DR2 captures this requirement by ensuring that these changes are directly propagated by the organization that owns the master data.

**DR3: Ensure Unambiguity of Master Data Across Organizational Boundaries.** To eliminate interpretive ambiguity during inter-organizational exchange, master data records must be governed by deterministic definitions that ensure semantic alignment. DR3 directly increases the reliability of master data across organizational boundaries (MDR1) and enables accountability for master data (MDR3) by ensuring that all parties operate under an identical, objective understanding of the data's meaning and context.

**DR4: Enforce Control over Master Data Usage.** Organizations are often hesitant to share master data due to a lack of clarity about usage scenarios. DR4 directly addresses sovereignty over master data lifecycle decisions (MDR2) and sets a baseline for the accountability of the master data exchange (MDR3). This enforcement is critical, as only the master data owner can ultimately guarantee its reliability and thus also account for it.

Based on derived DRs, we formulate our DPs following the guidelines of Chandra et al. [27] to keep our DPs actionable and effective. In addition to our DPs, we further detail their content by providing additional DFs. These DFs demonstrate how our DPs can be instantiated by concrete technical features [25].

**DP1: Provide Sovereign Identification Methods for Master Data Provenance.** Provide the system with sovereign identification methods for master data providers in order for users to unambiguously assign master data claims, given that master data provenance needs to be verified across organizational boundaries. DP1 increases verifiability of master data provenance (DR1) through a clear identification of the master data providers and ensures the unambiguity of master data records across organizations (DR3).

DF1: Decentralized Identifiers for Organizations (W3C Decentralized Identifiers)

DF2: Interoperable Trust Registries (Trust Frameworks, e.g., Gaia-X)

**DP2: Provide Cross-Organizational Semantic Models for Master Data.** Provide the system with cross-organizationally defined semantic models for master data in order for users to unambiguously interpret shared master data, given that internal semantic schemas vary across organizational boundaries. DP2 enables the direct propagation of master data status changes (DR2) and ensures the unambiguity of master data across organizational boundaries (DR3).

DF3: Standardized Credential Schemas (W3C Verifiable Credentials)

DF4: Credential Revocations and Status Updates (W3C Bitstring Status Lists)

**DP3: Provide Sovereign Master Data Verification Methods for Master Data Integrity.** Provide the system with sovereign data verification mechanisms in order for users to assess the timeliness and integrity of the master data at the time of use, given that master data frequently needs to be updated or revoked. DP3 directly ensures the verifiability of master data (DR1) and ensures the direct propagation of status changes (DR2), which would lead to a failed validation on outdated data.

DF5: Integrity Protection of Credentials (Cryptographic Signatures, e.g., Ed25519)

DF6: Selective Disclosure (Selective Disclosure for JSON Web Tokens)

**DP4: Bind Master Data Records to Enforceable Usage Policies.** Provide the system with machine-readable usage policies in order for users to exercise granular control over master data dissemination, given that organizations lose physical control over their data usage once it is shared across organizational boundaries. DP4 ensures control over master data usage (DR4) by empowering the provider to explicitly formulate master data usage conditions and leverage policy-enforcement technologies.

DF7: Enforceable Usage Policies (W3C Open Digital Rights Language)

**DP5: Ensure Accountability over Master Data Exchange.** Provide the system with hashed audit trails in order for users to provide evidence over their shared master data, given that the exchange of master data should be accountable. DP5 ensures the system verifies the occurrence and integrity of these exchanges without revealing the actual data content through hashed audit trails. Moreover, it ensures that all lifecycle changes are observed (DR2) and reinforces the requirement for sovereign control over master data usage (DR4). In cases where cryptographic accountability alone is insufficient for dispute resolution, such as conflicting master data claims across jurisdictions, governance frameworks and legally recognized registries must serve as complementary arbitration mechanisms.

DF8: Hashed Audit Trails (Merkle Trees)

## 6 Reference Architecture for Sovereign MDM in Data Spaces

To demonstrate the applicability of our derived DPs, we present a reference architecture instantiating them within the IDSA Data Space Reference Architecture Model 4.0 [29]. As illustrated in Fig. 3, the architecture comprises four layers and one vertical service layer. The Identity Layer establishes a decentralized root of trust by integrating SSI technologies and thus removing reliance on central identity providers. Building on this, the MDM Services Layer secures master data provenance (DP1) by permanently binding records to DIDs, while embedding them in VCs to guarantee integrity and enable revocation (DP3). The Data Space Layer serves as the interoperability bridge between organizations, hosting dataspace-wide credential schemas to bridge internal semantic models into a shared vocabulary (DP2). Data catalogs enable asset discovery, while participant agent services and contract negotiation automate policy agreement prior to transfer. The Governance Layer maintains trust lists, governance frameworks, and machine-readable usage policies that govern contract negotiation (DP4). Finally, the Vertical Services Layer provides cross-cutting functionalities, including onboarding services, identity provisioning, and brokerage services that generate tamper-proof audit trails for system-wide accountability (DP5).

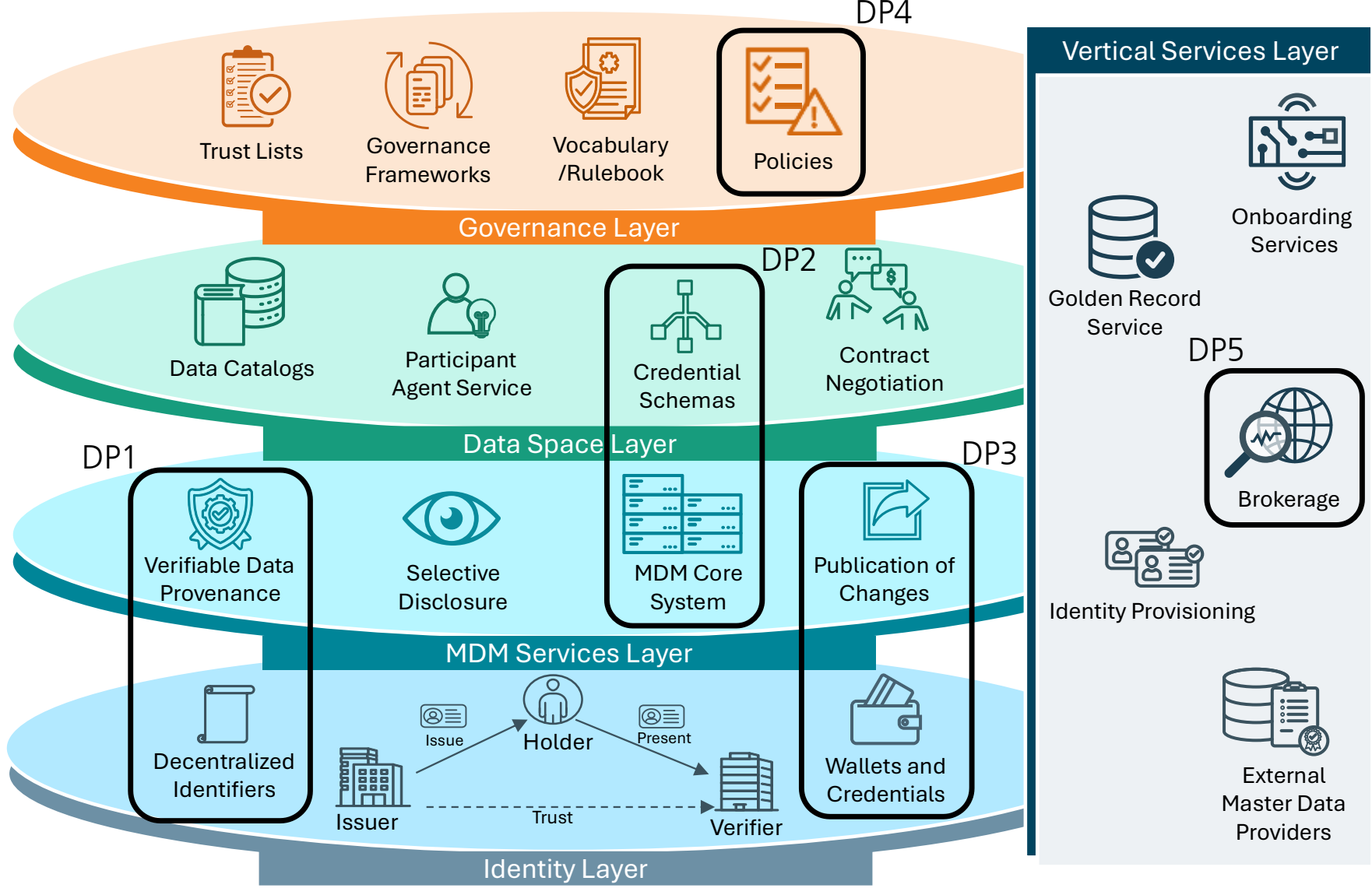

**Fig. 3.** Reference architecture for trustworthy MDM with SSI in data spaces.

## 7 Evaluation

Following the same purposive sampling strategy as in our problem identification interviews 4.3, we evaluated the nascent design theory through four expert interviews (E1–E4) targeting practitioners with and without SSI backgrounds across relevant domains. Overall, most experts confirmed the utility of the design theory, specifically highlighting the clear identification of organizations, semantic interoperability, and the integration of usage policies as helpful tools for building trust in cross-organizational master data exchange. However, the evaluation also revealed specific friction points regarding deficiency, redundancy, overload, and access to the design theory. Table 3 visualizes the demographics of our interview partners for evaluation.

**Table 3**. Interview demographics for DSR evaluation.

| E# | Role | Domain | Background in SSI? | Experience | Duration |
|---|---|---|---|---|---|
| E1 | Project Director | Automotive | yes | 7 years | 52 min |
| E2 | Innovation Manager | IT-Security | yes | 6 years | 31 min |
| E3 | Data Governance Manager | Energy | no | 10 years | 37 min |
| E4 | Digital Transformation Lead | Mobility | no | 20 years | 35 min |
| **Σ** | | | | | 154 min |

**Deficiency.** The interviews highlighted a gap regarding legal enforceability and cross-jurisdictional trust. E1 emphasized that “a trade register enjoys public faith” and that emerging EU regulations (Company Law, EU Business Wallet) will establish legally binding identity infrastructures by 2027/28, noting “if there are already legally higher-value sources [...] they should also be used” (E1). E2 similarly noted that “decentralizing master data means probably going across jurisdictions” as a “common pitfall” (E2).

**Redundancy.** Experts identified semantic overlaps between DP1 (“Ensure Unambiguous Interpretability of Master Data”) and DP5 (“Tamper-Proof Evidence of Master Data Exchange”). E2 observed that DP2 and DP3 could be combined, since technologies like VCs address both simultaneously (E2). E3 noted that “tamper-proof evidence [...] is essentially also master data provenance” (E3). We resolved this by shifting the focus of DP3 from general verifiability to master data integrity protection and DP5 to accountability of the master data exchange transaction.

**Overload.** Multiple experts noted that identification and semantic interoperability should not be combined within a single principle. E1 stated that "the identifiability of legal persons [...] has the potential to be explicitly listed" separately, while E3 reinforced that "semantic uniformity of master data" and provider identification serve distinct purposes. We addressed this by splitting the original DP1 into the current DP1 for identification of data provenance and DP2 for cross-organizational semantic interoperability. Regarding DP4, E1 distinguished between publicly available master data, where defining usage policies "means work [...] for hundreds of thousands of business partners," and sensitive master data, where such investment is justified. We nonetheless retained the enforceability claim in DP4 given its importance across interviews, despite its acknowledged technical limitations.

**Excess.** Experts noted that DP4 and DP5 might constitute technical excess for certain data types. E1 explicitly questioned whether "traceable data exchange" is "really necessary for public data" (E1), though ultimately confirmed relevance for most cases. E4 further suggested that continuous access to master data with local storage would potentially reduce the need for extensive audit trails (E4).

# 8 Discussion, Limitations, and Future Work

## 8.1 Theoretical Contribution and Practical Relevance

In this paper, we obtained a nascent design theory for trustworthy MDM with SSI. Our research contributes to the scientific knowledge base by addressing the intersection of traditional MDM and decentralized SSI architectures. Existing MDM literature predominantly relies on centralized data governance models, where central authorities enforce data quality. We challenge this assumption by presenting a nascent design theory that enables high-quality master data exchange in decentralized data sharing environments without a central "Golden Record.“ Furthermore, we broaden the application scope of SSI with our contribution. While SSI research typically restricts its focus to the provisioning of static identity claims or access control mechanisms [9–11], our artifact demonstrates how DIDs and VCs can also be leveraged to directly transfer master data records. By deriving DPs, we provide a blueprint for managing dynamic master data assets effectively in decentralized environments.

Practically, our design theory offers a concrete solution to resolve the tension between the operational efficiency necessary for master data exchange and digital sovereignty in emerging data ecosystems like Gaia-X or Catena-X. A primary obstacle for MDM identified in our expert interviews is an unclear status of master data assets, where organizations suffer from a lack of reliable updates on their partners' data changes. Our design theory addresses this by introducing mechanisms for the automated, real-time verification of master data validity and provenance. Moreover, by leveraging SSI-based trust anchors, organizations can automate the labor-intensive process of data provenance verification and data quality assurance, directly reducing the costs resulting from low master data quality. Moreover, our artifact provides a sovereign integration solution for MDM into ecosystems like Catena-X, effectively reducing dependencies on central services for master data quality improvements like the Cofinity-X GRS.

Our design is also aligned with relevant regulatory frameworks: the GDPR's data minimization and purpose limitation principles are directly supported by selective disclosure in DP2, the Data Act's requirements for non-discriminatory data access are addressed through sovereign, policy-controlled sharing in DP1 and DP3, and eIDAS 2.0's trust framework for legal entities provides a natural governance layer for anchoring organizational identities used in master data exchange.

### 8.2 Limitations and Future Work

It is important to note that decentralized MDM still needs to overcome a couple of adoption barriers. Centralized MDM registries offer well-understood governance structures, lower verification overhead for consumers, and mature tooling [35]. The advantage of our approach lies not in decentralization per se, but in its ability to preserve digital sovereignty while maintaining data quality assurance. This is a combination that centralized registries structurally cannot provide. SSI-based MDM also offers concrete advantages over alternative decentralized approaches such as federated APIs. It enables cryptographically verifiable provenance without requiring shared infrastructure, selective disclosure of master data attributes, and the ability to integrate with established trust frameworks like eIDAS 2.0 without introducing new intermediaries.

Beyond this technical trade-off, several adoption barriers require future investigation. Incentives for organizations to become credential issuers are non-trivial, as they need to bear the operational cost of credential issuance, while API-based alternatives may appear less costly in the short term. Thus, future work should explore governance models and incentive structures that make SSI-based issuance economically attractive. Also, migration from existing "Golden Record" systems and bootstrapping the critical mass of issuers and verifiers in ecosystems like Catena-X present further practical challenges that hybrid architectures and governance anchor institutions should address.

Beyond functional applicability, non-functional requirements also need to be considered for a successful adoption. Regarding scalability, credential issuance and verification are computationally lightweight operations, and the absence of a central bottleneck allows the architecture to scale horizontally across participants. Moreover, the overhead for key management represents a practical concern whereby organizations must maintain DID documents and credential lifecycle operations, which increases operational complexity compared to centralized alternatives.

This paper presents a first complete DSR cycle, focusing on derivation and formative, ex ante validation of the design theory. Our evaluation relied primarily on experts familiar with decentralized data ecosystems and SSI, which was necessary for validating technical feasibility but limits generalizability regarding broader business adoption. Future DSR iterations should therefore broaden stakeholder involvement to include MDM practitioners beyond the SSI domain, exploring adoption incentives, migration pathways, and government-backed trust anchors such as the EU Business Wallet as complementary governance infrastructure.

## 9 Conclusion

In this paper, we presented a nascent design theory for trustworthy MDM in decentralized data sharing environments, comprising three MDRs, five DPs, and eight DFs, instantiated through an integration into data space architectures. Our findings demonstrate that SSI provides a viable technical foundation for reliable, sovereign, and accountable cross-organizational master data exchange, without reliance on a central "Golden Records." Our design aligns with key regulatory frameworks, including

GDPR, the Data Act, and eIDAS 2.0, and offers a practical path toward reducing dependencies on central services in ecosystems like Catena-X. Future work should pursue subsequent DSR cycles to address these open challenges and validate the design theory in real-world deployments.

**Acknowledgments.** This work was supported by the Cluster of Excellence Cognitive Internet Technologies CCIT, which is funded by the Fraunhofer-Gesellschaft zur Förderung der angewandten Forschung e.V.

**Disclosure of Interests.** The authors declare that the research was conducted in the absence of any commercial or financial relationships that could represent a potential conflict of interest.

## References

1. Gaia-X European Association for Data and Cloud AISBL: Gaia-X, https://www.gaia-x.eu, last accessed 2026/01/22.
2. Catena-X Automotive Network e.V.: Catena-X: The First Open and Collaborative Data Ecosystem., https://catena-x.net/, last accessed 2026/01/22.
3. Otto, B., Ofner, M.: Strategic business requirements for master data management systems. In: Americas Conference on Information Systems (AMCIS). pp. 936–947 (2011).
4. Silvola, R., Jaaskelainen, O., Kropsu-Vehkapera, H., Haapasalo, H.: Managing one master data - Challenges and preconditions. Ind Manage Data Sys. 111, 146–162 (2011). https://doi.org/10.1108/02635571111099776.
5. Otto, B., Ebner, V., Hüner, K.: Measuring Master Data Quality: Findings from a Case Study. AMCIS 2010 Proceedings. (2010).
6. Keith, K., Seymour, L.: Precursors of Master Data Quality Issues across Enterprise Systems. UK Academy for Information Systems Conference Proceedings 2025. (2025).
7. Baghi, E., Otto, B., Oesterle, H.: Controlling Customer Master Data Quality: Findings from a Case Study. CONF-IRM 2013 Proceedings. (2013).
8. Roth, H., Mönch, S., Schäffer, T.: Towards Augmented MDM: Overview of Design and Function Areas – A Literature Review. AMCIS 2022 Proceedings. (2022).
9. Laatikainen, G., Kolehmainen, T., Abrahamsson, P.: Self-Sovereign Identity Ecosystems: Benefits and Challenges. 12th Scandinavian Conference on Information Systems. (2021).
10. Babel, M., Willburger, L., Lautenschlager, J., Völter, F., Guggenberger, T., Körner, M.-F., Sedlmeir, J., Strüker, J., Urbach, N.: Self-sovereign identity and digital wallets. Electron Markets. 35, 28 (2025). https://doi.org/10.1007/s12525-025-00772-0.
11. Schäfer, F., Rosen, J., Zimmermann, C., Wortmann, F.: Unleashing The Potential of Data Ecosystems: Establishing Digital Trust through Trust-Enhancing Technologies. European Conference on Information Systems (ECIS). (2023).
12. Vilminko-Heikkinen, R., Pekkola, S.: Changes in roles, responsibilities and ownership in organizing master data management. International Journal of Information Management. 47, 76–87 (2019). https://doi.org/10.1016/j.ijinfomgt.2018.12.017.
13. Hikmawati, S., Santosa, P.I., Hidayah, I.: Improving Data Quality and Data Governance Using Master Data Management: A Review. IJITEE. 5, 90 (2021). https://doi.org/10.22146/ijitee.66307.

14. Loshin, D.: Master Data Management. Elsevier Inc. (2009). https://doi.org/10.1016/B978-0-12-374225-4.X0001-X.
15. Cofinity-X: Golden Record Service, https://www.cofinity-x.com/golden-record, last accessed 2025/12/08.
16. Tobin, A., Reed, D., Windley, F.P.J., Foundation, S.: The Inevitable Rise of Self-Sovereign Identity. (2017).
17. Richter, D., Anke, J.: Self-sovereign identity: A conceptual framework and research agenda. Electron Markets. 36, 17 (2026). https://doi.org/10.1007/s12525-025-00867-8.
18. W3C: W3C Recommendation - Verifiable Credentials Data Model, https://www.w3.org/TR/vc-data-model-2.0/.
19. Laatikainen, G., Mustak, M., Hickman, N.: Self-sovereign identity adoption: Antecedents and potential outcomes. Technol. Soc. 82, (2025). https://doi.org/10.1016/j.techsoc.2025.102859.
20. Jeyakumar, I.H.J., Kubach, M.: A trust implementation model for cross-domain decentralized identity ecosystems: architecture, use case, and implementation. Procedia Computer Science. 254, 10–19 (2025). https://doi.org/10.1016/j.procs.2025.02.059.
21. Peffers, K., Tuunanen, T., Rothenberger, M.A., Chatterjee, S.: A design science research methodology for information systems research. Journal of Management Information Systems. 24, 45–77 (2007). https://doi.org/10.2753/MIS0742-1222240302.
22. vom Brocke, J., Winter, R., Hevner, A., Maedche, A.: Special Issue Editorial – Accumulation and Evolution of Design Knowledge in Design Science Research: A Journey Through Time and Space. Journal of the Association for Information Systems. 21, (2020). https://doi.org/10.17705/1jais.00611.
23. Hevner, A.R., March, S.T., Park, J., Ram, S.: Design Science in Information Systems Research1. MIS Quarterly. 28, 75–106 (2004). https://doi.org/10.2307/25148625.
24. Boell, S.K., Cecez-Kecmanovic, D.: A Hermeneutic Approach for Conducting Literature Reviews and Literature Searches. CAIS. 34, (2014). https://doi.org/10.17705/1CAIS.03412.
25. Herm, L.-V., Steinbach, T., Wanner, J., Janiesch, C.: A nascent design theory for explainable intelligent systems. Electron Markets. 32, 2185–2205 (2022). https://doi.org/10.1007/s12525-022-00606-3.
26. Jones, D., Gregor, S.: The Anatomy of a Design Theory. Journal of the Association for Information Systems. 8, (2007). https://doi.org/10.17705/1jais.00129.
27. Chandra, L., Seidel, S., Gregor, S.: Prescriptive Knowledge in IS Research: Conceptualizing Design Principles in Terms of Materiality, Action, and Boundary Conditions. In: Hawaii International Conference on System Sciences (HICSS). pp. 4039–4048. IEEE Computer Society, USA (2015). https://doi.org/10.1109/HICSS.2015.485.
28. Gregor, S., Kruse, L.C., Seidel, S.: Research Perspectives: The Anatomy of a Design Principle. Journal of the Association for Information Systems. 21, (2020). https://doi.org/10.17705/1jais.00649.
29. International Data Space Association: International Data Space Reference Architecture Model 4.0, https://internationaldataspaces.org/offers/reference-architecture/, last accessed 2026/01/26.

30. Venable, J., Pries-Heje, J., Baskerville, R.: FEDS: a Framework for Evaluation in Design Science Research. European Journal of Information Systems. 25, 77–89 (2016). https://doi.org/10.1057/ejis.2014.36.
31. Janiesch, C., Rosenkranz, C., Scholten, U.: An Information Systems Design Theory for Service Network Effects. Journal of the Association for Information Systems. 21, (2020). https://doi.org/10.17705/1jais.00642.
32. Raharjo, T., Abdurrahman, M.H., Yossy, E.H.: A model of critical success factors for master data management development projects using analytic hierarchy process (AHP): An insight from indonesia. In: International Conference on Management Science and Industrial Engineering. pp. 17–22. , Chiang Mai, Thailand (2023). https://doi.org/10.1145/3603955.3603959.
33. Silvola, R., Jaaskelainen, O., Kropsu-Vehkapera, H., Haapasalo, H.: Managing one master data – challenges and preconditions. Industrial Management & Data Systems. 111, 146–162 (2011). https://doi.org/10.1108/02635571111099776.
34. Buffenoir, E., Bourdon, I.: Reconciling Complex Organizations and Data Management: The Panopticon Paradigm. Pacific Asia Conference on Information Systems (PACIS). (2013).
35. Otto, B., Hüner, K.M., Österle, H.: Toward a functional reference model for master data quality management. Inf. Syst. e-Bus. Manage. 10, 395–425 (2012). https://doi.org/10.1007/s10257-011-0178-0.
36. S. K. Datta, T. Bokan, L. Resman: A Reference Architecture for Agricultural Data Spaces: Case Study from DIVINE Project. In: 2025 International Wireless Communications and Mobile Computing (IWCMC). pp. 8–13 (2025). https://doi.org/10.1109/IWCMC65282.2025.11059474.
37. Altendeitering, M., Guggenberger, T.M.: Data Quality Tools: Towards a Software Reference Architecture. Hawaii International Conference on System Sciences 2024 (HICSS-57). (2024).
38. Bader, S., Pullmann, J., Mader, C., Tramp, S., Quix, C., Mueller, A.W., Akyürek, H., Böckmann, M., Imbusch, B.T., Lipp, J., Geisler, S., Lange, C.: The International Data Spaces Information Model – An Ontology for Sovereign Exchange of Digital Content. Presented at the Lecture Notes in Computer Science (2020). https://doi.org/10.1007/978-3-030-62466-8_12.
39. Azkan, C., Möller, F., Ebel, M., Iqbal, T., Otto, B., Poeppelbuss, J.: Hunting the Treasure: Modeling Data Ecosystem Value Co-Creation. International Conference on Information Systems (ICIS). (2022).
40. Gelhaar, J., Otto, B.: Challenges in the Emergence of Data Ecosystems. Pacific Asia Conference on Information Systems (PACIS). (2020).
41. Barclay, I., Preece, A., Taylor, I., Radha, S.K., Nabrzyski, J.: Providing assurance and scrutability on shared data and machine learning models with verifiable credentials. Concurr. Comput. Pract. Exper. 35, (2023). https://doi.org/10.1002/cpe.6997.
42. Jarke, M., Otto, B., Ram, S.: Data Sovereignty and Data Space Ecosystems. Business & Information Systems Engineering. 61, 549–550 (2019).
43. Weber, K., Otto, B., Oesterle, H.: One Size Does Not Fit All---A Contingency Approach to Data Governance. ACM Journal of Data and Information Quality. 1, Article 4 (2009). https://doi.org/10.1145/1515693.1515696.